\documentclass[twoside]{article}
\usepackage{fleqn,espcrc2}

\usepackage{graphicx}

\newcommand{\AmS}{{\protect\the\textfont2
  A\kern-.1667em\lower.5ex\hbox{M}\kern-.125emS}}

\title{ An Exact Solution of BPS Junctions and Its Properties}

\author{ Kenji~Ito,
        Masashi~Naganuma, 
        Hodaka~Oda
        \thanks{
        The author gratefully acknowledges support from 
        the Iwanami Fujukai Foundation.}
        and
        Norisuke~Sakai
        \thanks{This work is supported in part 
        by Grant-in-Aid for Scientific Research from the Japan Ministry 
        of Education, Science and Culture for the Priority Area 291 and 707.}
\\ \vspace{0.3cm}
Department of Physics, Tokyo Institute of Technology, \\ 
        Oh-okayama, Meguro, Tokyo 152-8551, Japan
}
       
\begin{document}

\begin{abstract}
We have obtained an exact solution for the BPS domain wall junction 
 for a ${\cal N}=1$ supersymmetric theory in four dimensions and 
studied its properties. 
 The model is a simplified version of the ${\cal N}=2$ $SU(2)$ gauge 
theory with $N_f=1$ broken to ${\cal N}=1$ by the mass of the adjoint chiral 
superfield. 
We define mode equations and demonstrate explicitly that fermion and 
boson with the same mass have to come in pairs except massless modes. 
We work out explicitly massless Nambu-Goldstone (NG) 
modes on the BPS domain wall junction. 
We find that their wave functions extend along the wall to infinity 
(not localized) and are not normalizable. 
It is argued that this feature is a generic phenomenon of NG 
modes on domain wall junctions in the bulk flat space in any 
dimensions. 
NG fermions exhibit a chiral structure 
in accordance with unitary representations of $(1, 0)$ supersymmetry algebra 
where fermion and boson with 
the same mass come in pairs except massless modes which can appear singly. 
More detailed exposition of our results can be found in Refs.\cite{OINS},
\cite{INOS}. 
\vspace{1pc}
\end{abstract}

 \maketitle

\section{Introduction }

{}For sometime much attention has been paid to topological objects. 
Recently 
 an interesting idea has been advocated to regard 
 our world as a domain wall embedded in higher dimensional spacetime
\cite{ADD}. 
In this ``brane world" scenario, 
 our four-dimensional space-time on these topological objects 
 is embedded in higher dimensional space-time. 
Most of the particles in the standard model should be realized as 
modes localized on the wall. 
Phenomenological implications of the idea have been extensively studied 
from many aspects. 
Another fascinating possibility has also been proposed 
to consider walls in the bulk spacetime with a negative cosmological 
constant\cite{RS}. 

Supersymmetry has been useful to achieve stability of solitonic solutions 
such as domain walls. 
Domain walls in supersymmetric theories can saturate the Bogomol'nyi 
bound 
 and is called a 
 BPS state\cite{WittenOlive}. 
It has also been noted that these BPS states possess a topological charge 
which becomes a central charge $Z$ of the supersymmetry algebra 
\cite{AbrahamTownsend},\cite{DvaliShifman}. 
Thanks to the topological charge, these BPS states are guaranteed to be stable 
under arbitrary local fluctuations. 
The modes on the domain wall background 
have been worked out 
and are found to contain fermions and/or bosons localized on the wall 
in many cases\cite{RubakovShaposhnikov},\cite{ChibisovShifman}. 

By interpolating two discrete degenerate vacua in separate regions of space, 
we obtain a domain wall\cite{KSS},\cite{KSY}. 
These walls are typically codimension one and 
breaks half of the supersymmetry which 
is called a $1/2$ BPS state. 
If we have three or more discrete vacua in separate 
regions of space, 
segments of domain walls can meet at a one-dimensional junction 
and there arises a 
domain wall junction. 
The domain wall junction typically can preserve $1/4$ of supersymmetry 
and is called a $1/4$ BPS state. 

In this report we summarize our recent results on the first exact solution 
of BPS domain wall junction in a supersymmetric gauge 
theory\cite{OINS} and 
the detailed study of its properties\cite{INOS}.

\section{ BPS equations and $(1, 0)$ SUSY
 }

\subsection{ BPS equations}
If the translational invariance is broken as is the case for domain 
walls and/or junctions, the ${\cal N}=1$ superalgebra in general receives 
contributions from central charges 
\cite{AbrahamTownsend},\cite{DvaliShifman},\cite{OINS},\cite{INOS}. 
The anti-commutator between two left-handed supercharges has central 
charges 
$Z_k$,  $k=1,2,3$ 
\begin{equation}
\{ Q_\alpha, Q_\beta  \} =
2 i (\sigma^k \bar{\sigma}^0 )_\alpha{}^{\gamma} 
\epsilon_{\gamma \beta} Z_k. 
\label{qqantcom}
\end{equation}
The four dimensional indices 
are denoted by Greek letters $\mu, \nu =0, 1, 2, 3$ instead of 
roman letters $m, n $. 
The anti-commutator between left- and right-handed supercharges 
receives a contribution from central charges $Y_k, \ k=1,2,3$  
besides the energy-momentum 
four-vector $P^{\mu}, \ {\mu}=0, \cdots, 3$ of the system 
\begin{equation}
\{ Q_\alpha, \bar{Q}_{\dot{\alpha}} \} =
2 (\sigma^{\mu}_{\alpha \dot{\alpha}} P_{\mu} 
+ \sigma^k_{\alpha \dot{\alpha}} Y_k ) .
\label{qqantcomy}
\end{equation}
One may call $Z_k$ and $Y_k$ as $(1, 0)$ and $(1/2, 1/2)$ central charges 
in accordance with the transformation properties under the Lorentz group. 
Central charges, $Z_k$ and $Y_k$, come from the total divergence, 
and they are non-vanishing when there are nontrivial differences 
in asymptotic behavior of fields in different region of spatial infinity 
as is the case of domain walls and junctions\cite{GibbonsTownsend},\cite{CHT}. 
Therefore these charges are topological in the sense that they are 
determined completely by the boundary conditions at infinity. 
{}For instance, we can 
take a general Wess-Zumino model with 
an arbitrary number of chiral superfields $\Phi^i$, 
an arbitrary superpotential ${\cal W}$ 
 and an arbitrary K\"ahler potential $K(\Phi^i, \Phi^{* j})$ 
\begin{equation}
{\cal L}=\int d^2\theta d^2\bar{\theta} K(\Phi^i,\Phi^{* j}) 
 + \left[ \int d^2\theta {\cal W}(\Phi^i) 
+ \mbox{h.c.} \right],
\label{generalWZmodel}
\end{equation}
and compute the anticommutators (\ref{qqantcom}), 
(\ref{qqantcomy}) to find the central charges. 
The contributions to these central charges from bosonic components of 
chiral superfields are given 
by\cite{CHT}
\begin{equation}
Z_k = 2 \int d^3 x \, \partial_k {\cal W}^*(A^*),
\label{centralchargeZ}
\end{equation}
\begin{equation}
Y_k =i \epsilon^{knm}
\int d^3 x \, K_{i j^*}
\partial_n (A^{*j} \partial_m A^i), 
\label{centralchargeY}
\end{equation}
where $\epsilon^{123}=1$, and 
$A^i$ is the scalar component of the $i$-th chiral 
superfield $\Phi^i$ and  
$ K_{i j^*}=\partial^2 K(A^*, A) /\partial A^i\partial A^{*j}$ 
is the K\"ahler metric.

BPS domain wall is a $1/2$ BPS state\cite{DvaliShifman} 
and BPS domain wall junction is a $1/4$ BPS state
\cite{GibbonsTownsend},\cite{CHT}. 
To find the BPS equations satisfied by these BPS states, 
we consider a hermitian linear combination of operators 
$Q$ and $\bar{Q}$ with an arbitrary complex two-vector $\beta^\alpha$ 
and its complex conjugate $\bar{\beta}^{\dot{\alpha}} = (\beta^\alpha)^*$ 
as coefficients 
\begin{equation}
K= \beta^\alpha Q_\alpha + 
\bar{\beta}^{\dot{\alpha}} \bar{Q}_{\dot{\alpha}}. 
\end{equation}
We treat $\beta^\alpha$ as c-numbers rather than
the Grassmann numbers. 
Since $K$ is hermitian, 
the expectation value of the square of $K$ over any state 
is non-negative definite
\begin{equation}
\langle S | K^2 | S \rangle \ge 0.
\end{equation}
The field configuration of static junction must be at least two-dimensional. 
If we assume, for simplicity, that 
it depends on $x^1$, $x^2$ 
then we obtain $\langle Z_3 \rangle=\langle Y_1 \rangle
=\langle Y_2 \rangle = 0$ from Eqs.(\ref{centralchargeZ}) 
and (\ref{centralchargeY}), and 
the equality holds if and only if the linear combination of supercharges, 
$K$, is preserved by the state $ |S \rangle$
\begin{equation}
K \left\vert S \right\rangle= 0 . 
\end{equation}
In this case, the state $ |S \rangle$ saturates the energy bound and 
is called a BPS state. 
We find that there are two candidates 
for the saturation of the energy bound\cite{OINS}; 
\begin{equation}
H = H_{\rm I} \equiv
|\langle -i Z_1-Z_2 \rangle|
-\langle Y_3 \rangle, 
\end{equation}
when $
\bar{\beta}^{\dot{1}}=\beta^1
\frac{\langle i Z_1+Z_2 \rangle}{{|\langle i Z_1+Z_2 \rangle|}}$, and 
$\beta^2=\bar{\beta}^{\dot{2}}=0$,  
\begin{equation}
H = H_{\rm II} \equiv
|\langle i Z_1-Z_2 \rangle|
+\langle Y_3 \rangle, 
\end{equation}
when $
\beta^1=\bar{\beta}^{\dot{1}}=0$ and 
$\bar{\beta}^{\dot{2}}=\beta^2
\frac{\langle -i Z_1+Z_2 \rangle}{{|\langle -i Z_1+Z_2 \rangle|}}$.

In the case of $H_{\rm I}\ne H_{\rm II}$, 
the BPS bound becomes 
$
\langle H \rangle \ge 
{\rm max}\{ 
H_{\rm I} , H_{\rm II}
\} .
$
If $H_{\rm I} > H_{\rm II}$, 
then supersymmetry can only be preserved at $\langle H \rangle=H_{\rm I}$ 
and 
the only one combination of supercharges is  conserved 
\begin{equation}
\left(
Q_1 + 
\frac{\langle i Z_1+Z_2 \rangle}
{|\langle i Z_1+Z_2 \rangle|}
\bar{Q}_{\dot{1}}
\right)
| \langle H \rangle=H_{\rm I} \rangle =0.
\label{conseved_charge_1st}
\end{equation}
If $H_{\rm II}>H_{\rm I}$, 
then supersymmetry can only be preserved at 
$\langle H \rangle = H_{\rm II}$ and
the only one combination of supercharges is  conserved 
\begin{equation}
\left(
Q_2 + 
\frac{\langle -i Z_1+Z_2 \rangle}
{|\langle -i Z_1+Z_2 \rangle|}
\bar{Q}_{\dot{2}}
\right)
| \langle H \rangle=H_{\rm II} \rangle =0.
\label{conseved_charge_2st}
\end{equation}
In the case of $H_{\rm I}=H_{\rm II}$, 
two candidates of BPS bounds coincide 
 and BPS state conserves both of two supercharges, 
(\ref{conseved_charge_1st}) and (\ref{conseved_charge_2st}); 
this is a $1/2$ BPS state. 

{}For the general Wess-Zumino model in Eq.(\ref{generalWZmodel}), 
the condition of supercharge conservation (\ref{conseved_charge_1st}) 
for $H=H_{\rm I}$ applied to chiral superfield 
$\Phi^i=(A^i, \psi^i, F^i)$ gives 
after eliminating the auxiliary field $F^i$ 
\begin{equation}
2 
{\partial A^i \over \partial \bar{z}} 
=- \Omega_{+} F^i 
= \Omega_{+}K^{-1 i j^*} \frac{\partial {\cal W}^*}{ \partial A^{*j}}, 
\label{Be1}
\end{equation}
where $\Omega_{+}\equiv i \langle -i Z_1^*+Z_2^* \rangle/
|\langle -i Z_1^*+Z_2^* \rangle|$, and 
 complex coordinates $z=x^1+i x^2, \bar{z}=x^1-i x^2$, 
and the inverse of the K\"ahler metric $K^{-1 i j^*}$ 
are introduced. 
We can also consider gauge interactions where we should use covariant 
derivative instead of ordinary derivative. 
Moreover the same BPS condition (\ref{conseved_charge_1st}) applied to 
vector 
superfield in the Wess-Zumino gauge $V=(v_\mu, \lambda, D)$ gives 
after eliminating the auxiliary field $D$  
\begin{equation}
v_{12}=-D={1 \over 2} \sum_j A^{*j} e_j A^j, 
\label{Be1vector}
\end{equation}
and $ v_{03}=0, v_{01}=v_{31},  v_{23}=-v_{02}$, 
where 
$v_{\mu \nu}\equiv 
\partial_{\mu} v_{\nu} - \partial_{\nu} v_{\mu}$ and $e_j$ is the charge 
of the field $A^j$. 
Here we assume for simplicity the minimal 
kinetic term both for 
the chiral superfield $K_{i j^*} = \delta_{i j^*}$ 
and for the vector superfield. 

Similarly the condition of supercharge conservation 
(\ref{conseved_charge_2st}) 
for $H=H_{\rm II}$ applied to chiral superfield  in the Wess-Zumino model 
gives after eliminating the auxiliary field 
\begin{equation}
2 
{\partial A^i \over \partial z} 
=- \Omega_{-} F^i 
= \Omega_{-}K^{-1 i j^*} \frac{\partial {\cal W}^*}{ \partial A^{*j}}, 
\label{Be2}
\end{equation}
with 
$\Omega_{-}\equiv i\langle -i Z_1^*-Z_2^* \rangle/
{|\langle -i Z_1^*-Z_2^* \rangle|}$ .
If $U(1)$ gauge interaction is present, the derivative 
$\partial A^i/ \partial z$ should be replaced by the covariant derivative 
${\cal D}_z A^i=\frac{1}{2}({\cal D}_1-i {\cal D}_2) A^i$. 
 In this case 
the BPS condition applied to $U(1)$ vector 
superfield in the Wess-Zumino gauge 
 becomes
in the case of minimal kinetic terms
\begin{equation}
v_{12}=D=-{1 \over 2} \sum_j A^{*j} e_j A^j, 
\label{Be2vector}
\end{equation}
and $ v_{03}=0, v_{01}=-v_{31},  v_{23}=v_{02}$ . 

\subsection{The exact solution of BPS domain wall junction}

In ref.~\cite{OINS}, we have found an 
exact solution of BPS domain wall junction in a model motivated by 
the ${\cal N}=2$ supersymmetric $SU(2)$ gauge theory with one flavor 
 broken to ${\cal N}=1$ by the mass of the adjoint chiral superfield. 
This model has the following chiral superfields with the charge 
assignment for the $U(1)\times U(1)'$ gauge group 
\begin{equation}
\begin{array}{cccccccc}
      & {\cal M} & \tilde{\cal M} & {\cal D} & \tilde{\cal D} 
      & {\cal Q} & \tilde{\cal Q} & T \\
U(1)  &  0       &  0             &   1      &  -1            
      &  1       & -1             &   0 \\
U(1)' &  1       & -1             &   1      &  -1  
      &  0       &  0             & 0  ,
\end{array}
\end{equation}
interacting with a superpotential 
$$
{\cal W}=(T-\Lambda) {\cal M} \tilde{{\cal M}}
+ (T+\Lambda) {\cal D} \tilde{{\cal D}}
+ (T-m) {\cal Q} \tilde{{\cal Q}}
- h^2 T, 
$$
where parameters $\Lambda$ and $h$ can be made real positive and a 
 parameter $m$ is complex\cite{OINS}. 
In this model there are three discrete vacua, 
\begin{eqnarray}
&\!\!\!&\!\!\!
T=\Lambda, \, {\cal M}=\tilde{\cal M} = h, \,
{\cal Q}=\tilde{\cal Q}={\cal D}=\tilde{\cal D}=0,
 \nonumber \\
&\!\!\!&\!\!\!
T=m, \, {\cal Q}=\tilde{\cal Q}= h, \,
{\cal M}=\tilde{\cal M}={\cal D}=\tilde{\cal D}=0,  
 \nonumber \\
&\!\!\!&\!\!\!
T=-\Lambda, \, {\cal D}=\tilde{\cal D}= h, \,
{\cal Q}=\tilde{\cal Q}={\cal M}=\tilde{\cal M}=0, 
\nonumber
\end{eqnarray}
which are called vacuum 1, 2, and 3 with ${\cal W}_1=-h^2 \Lambda$,
${\cal W}_2=-h^2 m$ and ${\cal W}_3=h^2 \Lambda$ respectively. 
When $m=i\sqrt{3}\Lambda$, this model becomes $Z_3$ symmetric. 
Thus three half walls are expected to connect at the junction 
with relative angles of $2\pi/3$.
{}For definiteness, we specify the boundary condition where the 
wall $1$ extends along the negative $x^2$ axis separating the vacuum 
$1$ ($x^1>0$) and $3$ ($x^1<0$). 
If we have only the wall $1$, we obtain the central charge $Z_k$ 
(vanishing $Y_k$) and find 
the two conserved supercharges 
from Eqs.(\ref{conseved_charge_1st}) 
and (\ref{conseved_charge_2st}) 
as 
\begin{eqnarray}
 Q^{(1)} 
 &\!\!\!=&\!\!\! 
\frac{1}{\sqrt{2}}({\rm e}^{-i{\pi \over 4}} Q_2 
+ {\rm e}^{i{\pi \over 4}} \overline{Q}_{\dot{2}}),
\\
  Q^{(2)}
 &\!\!\!=&\!\!\! 
\frac{1}{\sqrt{2}}({\rm e}^{i{\pi \over 4}}Q_1 
+{\rm e}^{-i{\pi \over 4}} \overline{Q}_{\dot{1}}). 
\end{eqnarray}
The other two walls have also two conserved supercharges : 
 at wall 2, 
\begin{equation}
 Q^{(3)} = \frac{1}{\sqrt{2}} 
                            ({\rm e}^{-i{\pi \over 12}} Q_1 
                          + {\rm e}^{i{\pi \over 12}} \overline{Q}_{\dot{1}})
\end{equation}
besides $Q^{(1)}$, 
 and 
at wall 3, 
\begin{equation}
          Q^{(4)} = \frac{1}{\sqrt{2}}
                   ({\rm e}^{-i{5\pi \over 12}} Q_1 
                  + {\rm e}^{i{5\pi \over 12}}\overline{Q}_{\dot{1}}), 
\end{equation}
besides $Q^{(1)}$. 
When these three half walls coexist, we can have only one common conserved 
supercharge $Q^{(1)}$. 
In fact we find that the domain wall junction configuration conserves 
precisely this single combination of supercharges, 
even though it has also another central charge $Y_k$ contributing. 
Correspondingly we obtain the BPS equations (\ref{Be2}) and (\ref{Be2vector}) 
for $H=H_{{\rm II}}$ with $\Omega_{-}=-1$. 
The BPS equations (\ref{Be2vector}) for the vector superfield can be trivially 
satisfied  by $v_{\mu}=0$ and $D=0$. 
The BPS equations (\ref{Be2}) for chiral superfields become in this case 
\begin{equation}
2 \frac{ \partial A^i }{ \partial z}
=  -\frac{\partial {\cal W}^*}{\partial A^{*i}} ,
\label{BPSeqs}
\end{equation}
assuming the minimal kinetic term. 
The solution for these BPS equations is given by\cite{OINS},
\begin{eqnarray}
{\cal M}(z, \bar{z})
&\!\!\! = &\!\!\! 
\tilde{\cal M}(z, \bar{z})=
\frac{\sqrt2 \Lambda s}{s+t+u}, \nonumber \\
{\cal D}(z, \bar{z})
&\!\!\! = &\!\!\! 
\tilde{\cal D}(z, \bar{z})=
\frac{\sqrt2 \Lambda t}{s+t+u}, \nonumber \\
{\cal Q}(z, \bar{z})
&\!\!\! = &\!\!\! 
\tilde{\cal Q}(z, \bar{z})=
\frac{\sqrt2 \Lambda u}{s+t+u},\label{exact solution} \\
T(z, \bar{z})
&\!\!\! = &\!\!\! 
\frac{2 \Lambda}{\sqrt{3}}
\frac{
e^{-i \frac{1}{6} \pi} s
+ e^{-i \frac{5}{6} \pi} t
+ e^{i \frac{1}{2} \pi} u
}{s+t+u}+\frac{i}{\sqrt3}\Lambda, \nonumber 
\end{eqnarray}
where $s=\exp \left[\frac{2}{\sqrt{3}} \Lambda 
{\rm Re} \left( e^{i \frac{1}{6} \pi} z \right)\right]$,
$t=\exp \left[ \frac{2}{\sqrt{3}} \Lambda \right.$ 
${\rm Re} \left. \left( e^{i \frac{5}{6} \pi} z \right)\right]$,
and $u=\exp \left[ \frac{2}{\sqrt{3}} \Lambda 
{\rm Re} \left( e^{-i \frac{1}{2} \pi} z \right)\right]$.

This model is motivated by the softly broken ${\cal N}=2$ $SU(2)$ 
gauge theory with one flavor. 
However, we can simplify the model without spoiling the solvability 
to obtain a Wess-Zumino model consisting of purely chiral superfields 
by the following procedure. 
The vector superfields actually serve to constrain chiral superfields 
to have the identical magnitude pairwise through $D=0$ 
to satisfy the BPS equation (\ref{Be2vector}) for vector superfields: 
$|\tilde{\cal M}|=|{\cal M}|, |\tilde{\cal D}|=|{\cal D}|, 
|\tilde{\cal Q}|=|{\cal Q}|$. 
Therefore we can eliminate the vector superfields and reduce the number of 
chiral superfields by identifying pairwise 
$\tilde{\cal M}={\cal M}, \tilde{\cal D}={\cal D}, 
\tilde{\cal Q}={\cal Q}$ \cite{INOS},\cite{ShifmanVeldhuis}. 

\subsection{Unitary representations of $(1, 0)$ supersymmetry algebra}
\label{sc:unitaryrep}
Let us examine states on the background of a domain wall junction from 
the point of view of surviving symmetry. 
In the case of the BPS states satisfying the BPS equation (\ref{Be2}) 
corresponding to $H=H_{{\rm II}}$, we 
have only one surviving supersymmetry charge $Q^{(1)}$, 
two translation generators $H, P^3$, and one Lorentz generator $J^{0 3}$, 
out of the ${\cal N}=1$ four dimensional super Poincar\'{e} generators. 
Since we are interested in excitation modes on the background of the 
domain wall junction, we define the hamiltonian $H'=H-\langle H \rangle$ 
measured from the energy $\langle H \rangle$ of the background configuration. 
By projecting from the supersymmetry algebra (\ref{qqantcom}), 
(\ref{qqantcomy}) with central charges in four dimensions, 
we immediately find 
\begin{equation}
\left(Q^{(1)}\right)^2=H' -P^3 . 
\end{equation}
We also obtain the Poincar\'{e} algebra in $1+ 1$ dimensions 
$$
[J^{03}, Q^{(1)}]= {i \over 2} Q^{(1)}, \, 
[J^{03}, H'\mp P^3]= i (H'\mp P^3). 
$$
Other commutation relations vanishes trivially. Together they form 
 the $(1, 0)$ supersymmetry algebra on the domain wall 
junction as anticipated\cite{GibbonsTownsend}. 

To obtain unitary representations, we can diagonalize $H'$ and $P^3$ 
\begin{equation}
H'\vert E, p^3 \rangle =E \vert E, p^3 \rangle, \, 
P^3 \vert E, p^3 \rangle =p^3 \vert E, p^3 \rangle, 
\end{equation}
with $E \ge |p^3|$ and combine them by means of $Q^{(1)}$. 
If $E-p^3 > 0$, we can construct bosonic state from fermionic state and vice 
versa by operating $Q^{(1)}$ on the states 
$( \vert B \rangle, \vert F \rangle )$, 
$\vert B \rangle =Q^{(1)} \vert F \rangle/\sqrt{E-p^3}$ and  
$\vert F \rangle =Q^{(1)} \vert B \rangle/\sqrt{E-p^3}$, 
obtaining a doublet representation.
If $E-p^3 = 0$, operating by 
 $Q^{(1)}$ on the state gives an unphysical zero norm state 
\begin{equation}
\left|Q^{(1)}\vert E, p^3 \rangle\right|^2 
=\langle E, p^3 \vert H'-P^3 \vert E, p^3 \rangle 
=0.
\end{equation}
Then the massless right-moving state $\vert E, p^3=E \rangle$ 
is  a singlet representation. 
This singlet state can either be boson or fermion. 
Thus we find that there are only two types of representations of the 
 $(1, 0)$ supersymmetry algebra, doublet and singlet. 
We also find that massive modes should appear in pairs of boson and fermion, 
whereas the massless right-moving mode can appear singly without 
accompanying a state with opposite statistics. 
This provides an interesting possibility of a chiral structure for 
fermions. 

If another BPS equation (\ref{Be1}) corresponding to $H=H_{\rm I}$ is 
satisfied instead of Eq. (\ref{Be2}), we have $(0, 1)$ 
supersymmetry and the left-moving massless states can appear as 
singlets.

\section{Nambu-Goldstone and other modes on the junction}

\subsection{Mode equation on the junction}

Since the vector superfields have no nontrivial field configurations, 
Nambu-Goldstone modes have no component of vector superfield. 
Moreover we can replace our model, if we wish, by another 
model with purely chiral superfields without spoiling the 
essential features including the solvability. 
Consequently we shall neglect 
vector superfields and consider the general Wess-Zumino model 
in Eq.(\ref{generalWZmodel}) in the following. 
{}For simplicity we assume the minimal kinetic term 
here $K_{ij^*}=\delta_{ij^*}$ .

Let us consider 
quantum fluctuations $A'^i, \psi^i$ 
around a classical solution $A^i_{\rm cl}$ which satisfies the BPS equations 
 (\ref{Be1}) and (\ref{Be1vector}) for $H=H_{\rm I}$ 
or (\ref{Be2}) and (\ref{Be2vector}) for $H=H_{\rm II}$. 
The linearized equation for fermions is given by 
\begin{equation}
-i \bar \sigma^\mu \partial_\mu \psi^i - 
{\partial^2 {\cal W}^* \over \partial A^{*i}_{\rm cl}\partial A^{*j}_{\rm cl}}
\bar \psi^j = 0
\label{eq:lin-fermion1}
\end{equation}
\begin{equation}
-i \sigma^\mu \partial_\mu \bar \psi^i - 
{\partial^2 {\cal W} \over \partial A^{i}_{\rm cl}\partial A^{j}_{\rm cl}}
 \psi^j = 0 .
\label{eq:lin-fermion2}
\end{equation}
To separate variables for fermion equations, it is more convenient to use 
a gamma matrix representation where direct product structure of $2 \times 2$ 
matrices for $(x^0,x^3)$ and $(x^1, x^2)$ space is manifest. 
Transforming from such a representation to the Weyl representation 
which we are using, we can define the fermionic 
modes $\psi^i_{n \alpha}, \bar \psi^{i \dot \beta}_n$ 
combining components of left-handed and right-handed spinors 
by means of the following operators 
\begin{equation}
{\cal O}_1{}^i{}_j \!\! \equiv \!\!
\left[\matrix{
- {\partial^2 {\cal W}^* \over \partial A^{*i}_{\rm cl}
\partial A^{*j}_{\rm cl}}
&
-i\left(-\partial_1+i\partial_2\right) \delta^i_j \cr
-i\left(\partial_1+i\partial_2\right) \delta^i_j 
&
- {\partial^2 {\cal W} \over \partial A^{i}_{\rm cl}\partial A^{j}_{\rm cl}}
}\right]
\label{eq:fermion-mode-weyl1}
\end{equation}
\begin{equation}
{\cal O}_2{}^i{}_j \!\! \equiv \!\! 
\left[\matrix{
- {\partial^2 {\cal W} \over \partial A^{i}_{\rm cl}\partial A^{j}_{\rm cl}}
&
-i\left(\partial_1-i\partial_2\right) \delta^i_j \cr
-i\left(-\partial_1-i\partial_2\right) \delta^i_j 
&
- {\partial^2 {\cal W}^* \over \partial A^{*i}_{\rm cl} 
\partial A^{*j}_{\rm cl}}
}\right]
\label{eq:fermion-mode-weyl2}
\end{equation}
\begin{equation}
{\cal O}_1{}^i{}_j 
\left[\matrix{\bar \psi^{j \dot 1}_n \cr \psi^j_{n 2} \cr}\right]
 = -i m_n^{(1)} 
\left[\matrix{ \psi^i_{n 1} \cr \bar \psi^{i \dot 2}_n \cr}\right]
\label{eq:fermion-mode-weyl3}
\end{equation}
\begin{equation}
{\cal O}_2{}^i{}_j 
\left[\matrix{\psi^j_{n 1} \cr \bar \psi^{j \dot 2}_n \cr}\right]
 = i m_n^{(2)} 
\left[\matrix{\bar \psi^{i \dot 1}_n \cr \psi^i_{n 2} \cr}\right] ,
\label{eq:fermion-mode-weyl4}
\end{equation}
where the mass eigenvalues $m_n^{(1)}, m_n^{(2)}$ are real. 
Please note a peculiar combination of left- and right-handed 
spinor components to define eigenfunctions. 
We can expand $\psi^i$ in terms of these mode functions 
\begin{equation}
 \psi^i_\alpha (x^0, x^1, x^2, x^3)\!\!=\!\!\sum_n 
\left(\matrix{b_n(x^0, x^3) \psi^i_{n1}(x^1, x^2)
\cr 
c_n(x^0, x^3) \psi^i_{n2}(x^1, x^2)\cr}\right)
\end{equation}
Since $\psi(x^0,x^1,x^2,x^3)$ is a Majorana spinor, the coefficient 
fermionic fields $b_n, c_n$ are real. 
The linearized equations (\ref{eq:lin-fermion1}) (\ref{eq:lin-fermion2}) 
for the fermion gives a Dirac equation 
in $1+1$ dimensions for the coefficient fermionic fields $\phi_n=(c_n, i b_n)$ 
with two mass parameters $m_n^{(1)}, m_n^{(2)}$ 
\begin{equation}
 \left[-i \gamma^a\partial_a  
-m_n^{(1)}{1 + \rho_3 \over 2} 
-m_n^{(2)}{1 - \rho_3 \over 2} \right]\phi_n
 =0 ,
\label{eq:dirac1+1}
\end{equation}
where we use Pauli matrices $\rho_a,  \ a=1,2,3$ to construct the 
$2 \times 2$ gamma matrices in $1+1$ dimensions 
$(\gamma^0, \gamma^1)=(\rho_1, i\rho_2)$. 
Since we have a Majorana spinor in $1+1$ dimensions which does not allow 
 chiral rotations, we have 
two distinct real mass parameters $m_n^{(1)}, m_n^{(2)}$. 

Similarly 
we retain the part of the Lagrangian quadratic in fluctuations 
and eliminate the auxiliary fields $F^i$ to obtain the linearized 
equation for the scalar fluctuations, $A'^i = A^i-A^i_{\rm cl}$,  
\begin{eqnarray}
&\!\!\!-&\!\!\!\partial_\mu \partial^\mu A'^{*i}
+{\partial^2 {\cal W} \over \partial A^{i}_{\rm cl}\partial A^{k}_{\rm cl}}
{\partial^2 {\cal W}^* \over \partial A^{*k}_{\rm cl}\partial A^{*j}_{\rm cl}}
 A'^{*j} \nonumber \\
&\!\!\!+&\!\!\!{\partial^3 {\cal W} \over 
\partial A^{i}_{\rm cl}\partial A^{k}_{\rm cl}\partial A^{j}_{\rm cl}}
{\partial {\cal W}^* \over \partial A^{*k}_{\rm cl}}
 A'^{j}
 = 0 .
\end{eqnarray}
In order to separate variables 
in $x^0, x^3$ and $x^1, x^2$ we have 
to define mode equations on the background which has a nontrivial 
dependence in two dimensions, $x^1, x^2$. 
The bosonic modes $ A'^{i}_n(x^1, x^2)$ can easily be defined in terms of 
a differential operator ${\cal O}_B$ in $x^1, x^2$ space 
\begin{equation}
{\cal O}_B{}^i{}_j 
\left[\matrix{
A'^{*j}_n \cr 
 A'^{j}_n
\cr}\right]
 = M_n^2  
\left[\matrix{
A'^{*i}_n \cr 
 A'^{i}_n
\cr}\right] ,
\end{equation}
where the eigenvalue $M_n^2$ has to be real from Majorana condition. 
The quantum fluctuation for scalar can be expanded in terms of these 
mode functions to obtain a real scalar field equation with the mass $M_n$ 
for the coefficient bosonic field $a_n(x^0, x^3)$, 
\begin{equation}
 A'^{i}(x^0, x^1, x^2, x^3)=\sum_n a_n(x^0, x^3) A'^i_n(x^1, x^2)
\end{equation}
\begin{equation}
 \left(\partial_0^2 -\partial_3^2 + M_n^2 \right) a_n(x^0, x^3) =0 .
\end{equation}

To relate the mass eigenvalues of fermions and bosons, 
let us multiply two differential operators for fermions 
${\cal O}_2 $ to ${\cal O}_1$. 
In this ordering, we can use the BPS equation (\ref{Be2}) corresponding 
to $H=H_{{\rm II}}$ to find the differential operator for bosons ${\cal O}_B$ 
\begin{equation}
{\cal O}_2^i{}_k 
{\cal O}_1^k{}_j 
 = U
{\cal O}_B^i{}_j 
U^{-1},
\end{equation}
\begin{equation}
U=
 \left[\matrix{{\rm e}^{i{\pi \over 4}}\Omega_-^{1\over 2} & 0\cr 
               0 & {\rm e}^{-i{\pi \over 4}}\Omega_-^{-{1\over 2}}\cr}\right]. 
\end{equation}
Therefore the BPS equation (\ref{Be2}) corresponding 
to $H=H_{{\rm II}}$ guarantees that the existence of a solution 
$\bar \psi_n^{i\dot 1}, \psi_{n 2}^i$ 
of fermionic mode equations implies the existence of a solution of 
bosonic mode equations with the mass squared $M_n^2=m_n^{(1)}m_n^{(2)}$ 
\begin{equation}
A_n'^{*i}
={\rm e}^{-i{\pi \over 4}}\Omega_-^{-{1\over 2}}\bar \psi_n^{i\dot 1} ,
\qquad 
A_n'^{i}={\rm e}^{i{\pi \over 4}}\Omega_-^{{1\over 2}} \psi_{n2}^{i} . 
\end{equation}

If another 
BPS equation (\ref{Be1}) corresponding to $H=H_{\rm I}$ is valid, 
operator multiplication with different ordering gives the 
same bosonic operator whose rows and columns are interchanged 
\begin{equation}
{\cal O}_1^i{}_k 
{\cal O}_2^k{}_j 
 =U'
{\cal O}_B^i{}_j 
U'^{-1},
\end{equation}
\begin{equation}
U'=
  \left[\matrix{0 & {\rm e}^{i{\pi \over 4}}\Omega_+^{-{1\over 2}} \cr 
                 -{\rm e}^{-i{\pi \over 4}}\Omega_+^{1\over 2} & 0 \cr}\right].
\end{equation}
Therefore the BPS equation (\ref{Be1}) corresponding 
to $H=H_{\rm I}$ guarantees that the existence of a solution 
$\bar \psi_n^{i\dot 2}, \psi_{n1}^{i}$ 
of fermionic mode equations implies the existence of a solution of 
bosonic mode equations with the mass squared $M_n^2=m_n^{(1)}m_n^{(2)}$  
\begin{equation}
A_n'^{*i}
=-{\rm e}^{i{\pi \over 4}}\Omega_+^{-{1\over 2}}\bar \psi_n^{i\dot 2} , 
\qquad 
A_n'^{i}={\rm e}^{-i{\pi \over 4}}\Omega_+^{{1\over 2}} \psi_{n1}^{i} . 
\end{equation}

Therefore we find that all massive states come in pairs of boson and 
fermion with the same mass squared $M_n^2=m_n^{(1)}m_n^{(2)}$ 
in accordance with the result of the unitary representation of 
the $(1, 0)$ supersymmetry 
algebra. 


\subsection{Nambu-Goldstone modes}

Since we are usually most interested in a low energy effective field theory, 
we wish to study massless modes here. 
If global continuous symmetries are broken spontaneously, 
there occur associated massless modes which are called the Nambu-Goldstone 
modes. 
To find the wave functions of the Nambu-Goldstone modes, we perform 
the associated global transformations and evaluate 
the transformed configuration by substituting the classical field. 
For supersymmetry we obtain nontrivial wave function by substituting 
the classical field $A^i_{\rm cl}(x^1,x^2)$ and $F^i_{\rm cl}(x^1,x^2)$ 
to the transformation of fermions by a Grassmann parameter $\xi$, since 
classical field configuration of fermion vanishes $\psi^i_{\rm cl}=0$ 
\begin{equation}
\delta_{\xi} \psi^i = i\sqrt{2}\sigma^\mu\bar{\xi}\partial_\mu A^i_{\rm cl} 
+ \sqrt{2} \xi F^i_{\rm cl} .
\end{equation}
If the BPS equation (\ref{Be2}) for the junction background is valid, 
we obtain 
\begin{eqnarray}
&\!\!\!&\!\!\!\delta_{\xi} \psi^i =  \\
&\!\!\!&\!\!\!
\sqrt{2}\left[ 
(i\sigma^1\bar{\xi}-\Omega^*_{-}\xi)\partial_1 A^i_{\rm cl}
+(i\sigma^2\bar{\xi}+i\Omega^*_{-}\xi)\partial_2 A^i_{\rm cl}
\right] \nonumber
\end{eqnarray}
We see that there is one conserved direction in the Grassmann parameter: 
\begin{equation}
i\sigma^1\bar{\xi}\, =\, \Omega^*_{-}\xi \,\,\, \mbox{and} \,\,\, 
\sigma^2\bar{\xi}\, =\, -\Omega^*_{-}\xi .
\end{equation}
The other three real Grassmann parameters $\xi$ correspond to 
broken supercharges. 
{}For our exact solution, for instance, we find it convenient to choose 
the three broken supercharges as the following real supercharges 
\begin{equation}
  Q_{\rm I}= \frac{1}{\sqrt{2}}
              (e^{i\pi/4}Q_2 + e^{-i\pi/4}\bar{Q}_{\dot{2}}), 
\label{eq:brokencharges1}
\end{equation}
\begin{equation}
  Q_{\rm II}= \frac{1}{\sqrt{2}}
              (e^{-i\pi/4}Q_1 + e^{i\pi/4}\bar{Q}_{\dot{1}}), 
\end{equation}
\begin{equation}
  Q_{\rm III}= \frac{1}{\sqrt{2}}
              (e^{i\pi/4}Q_1 + e^{-i\pi/4}\bar{Q}_{\dot{1}}).
\label{eq:brokencharges}
\end{equation} 
Then the corresponding massless mode functions are given by 
\begin{eqnarray}
  \psi_0^{{\rm (I)}i}(x^1,x^2) 
    &\!\!\!=&\!\!\!
    \left( 
    \begin{array}{c}       
   4\partial_z A_{\rm cl}^i(x^1,x^2) e^{-i\pi/4} \\
       0 
    \end{array}
    \right),    
  \label{NG1} \\
  \psi_0^{{\rm (II)}i}(x^1,x^2)
    &\!\!\!=&\!\!\!  
    \left( 
    \begin{array}{c}
       0 \\
   2\partial_1 A_{\rm cl}^i(x^1,x^2)e^{i\pi/4} 
    \end{array}
    \right),
  \label{NG2} \\
  \psi_0^{{\rm (III)}i}(x^1,x^2)
    &\!\!\!=&\!\!\!  
    \left( 
    \begin{array}{c}
       0 \\ 
       2\partial_2 A_{\rm cl}^i(x^1,x^2) e^{i\pi/4} 
    \end{array}
    \right).
  \label{NG3} 
\end{eqnarray}
Since the transformation parameter should correspond to the Nambu-Goldstone 
field with zero momentum and energy, the three transformation parameters $\xi$ 
should be promoted to three real fermionic fields in $x^0, x^3$ space, 
 $b_0^{({\rm I})}(x^0, x^3), c_0^{({\rm II})}(x^0, x^3)$, and 
$c_0^{({\rm III})}(x^0, x^3)$, 
to obtain the Nambu-Goldstone component of the mode expansion 
\begin{eqnarray}
&\!\!\!&\!\!\!  \psi^{i}(x^0, x^1,x^2, x^3) 
 = 
b_0^{({\rm I})}(x^0, x^3)   \psi_0^{{\rm ({\rm I})}i}(x^1,x^2) 
\nonumber \\
&\!\!\!&\!\!\!
+ c_0^{({\rm II})}(x^0, x^3)   \psi_0^{{\rm ({\rm II})}i}(x^1,x^2)  
\nonumber \\
&\!\!\!&\!\!\!
+ 
 c_0^{({\rm III})}(x^0, x^3) 
  \psi_0^{{\rm ({\rm III})}i}(x^1,x^2) 
\nonumber \\
&\!\!\!&\!\!\!
+ \sum_{n > 0} 
\left(\matrix{b_n(x^0, x^3) \psi^i_{n1}(x^1, x^2)
\cr 
c_n(x^0, x^3) \psi^i_{n2}(x^1, x^2)\cr}\right).
\end{eqnarray}
We have explicitly displayed three massless Nambu-Goldstone fermion components 
 distinguishing from the massive ones ($n>0$). 
The Dirac equation for the coefficient fermionic fields (\ref{eq:dirac1+1}) 
shows that  $b_0^{({\rm I})}(x^0-x^3)$ is a right-moving massless mode, 
and $c_0^{({\rm II})}(x^0+x^3)$, and 
$c_0^{({\rm III})}(x^0+x^3)$ are left-moving modes. 

\begin{figure}[htbp]
\begin{eqnarray*}
\begin{array}{cc}
\includegraphics[width=8pc]{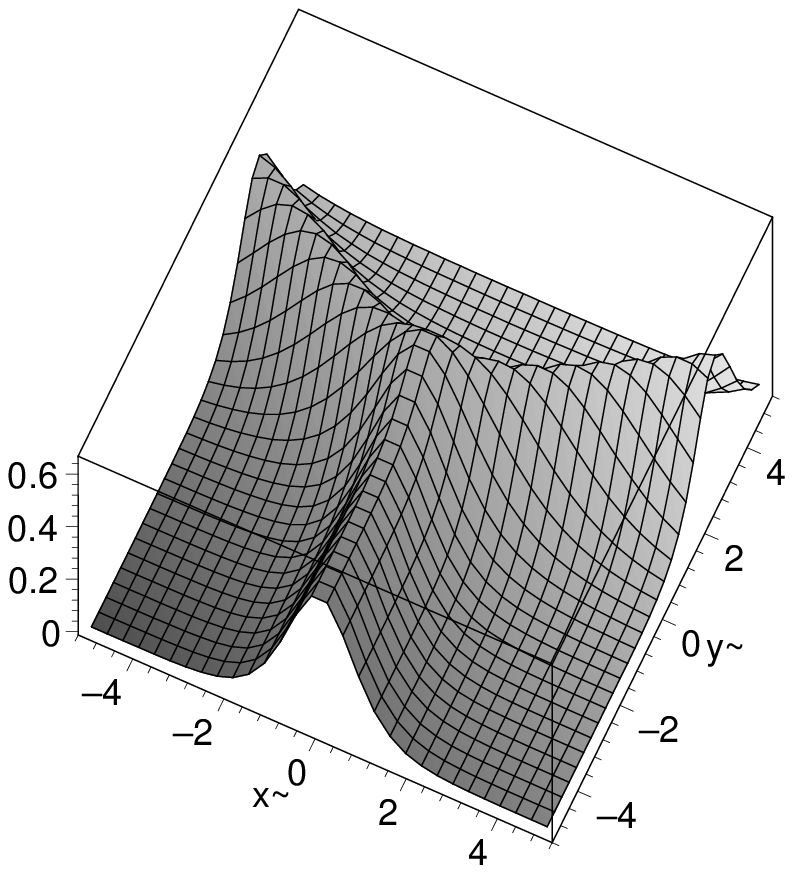} & 
\includegraphics[width=8pc]{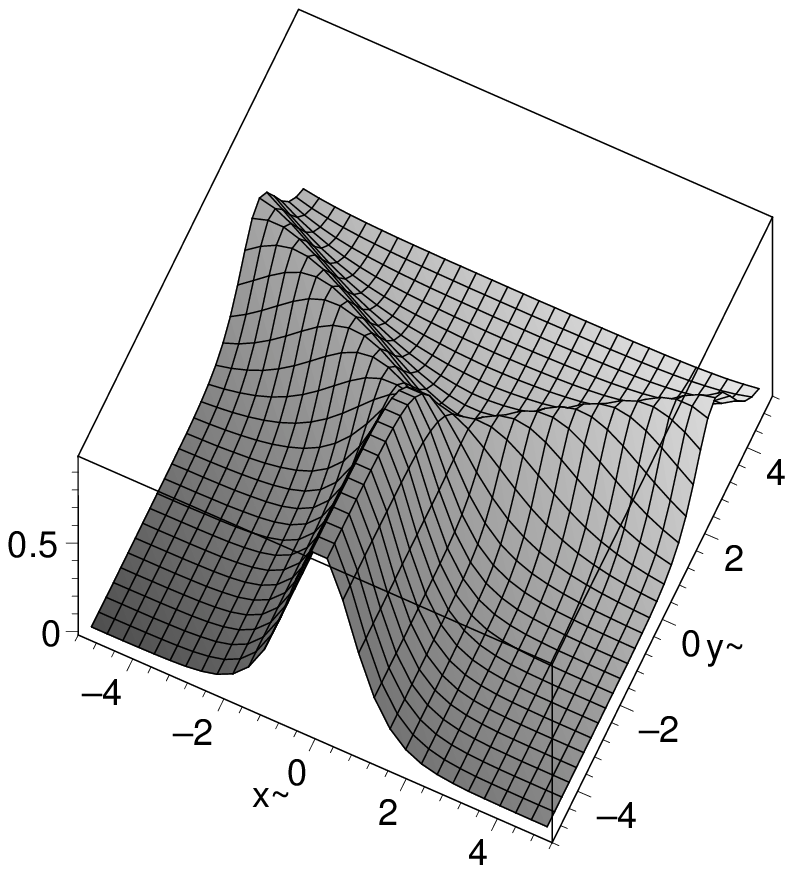} \\
\includegraphics[width=8pc]{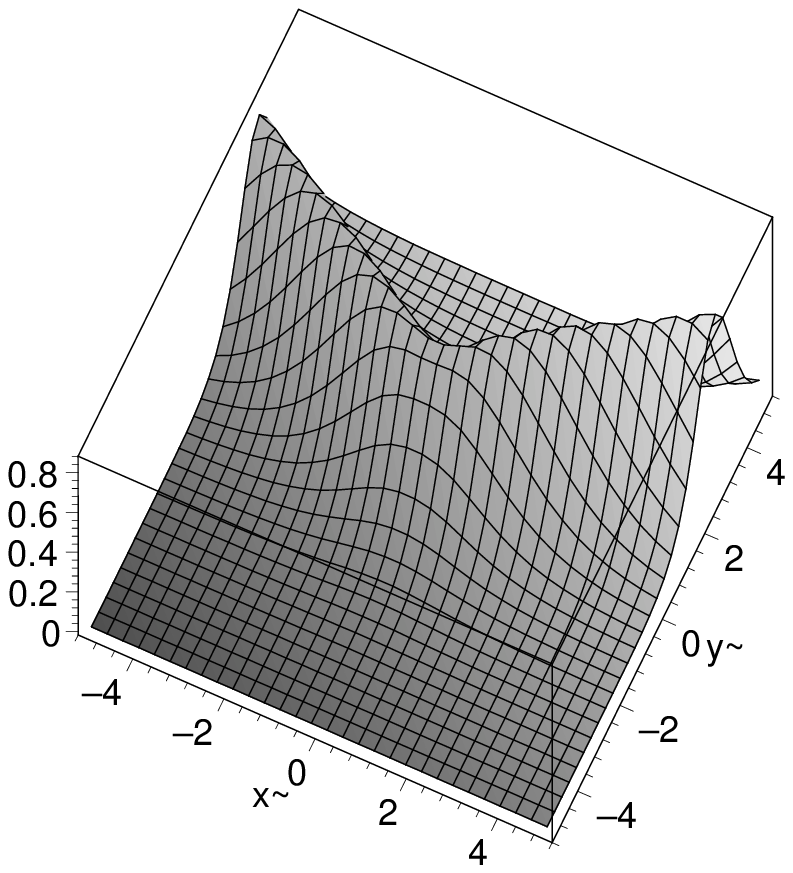} &      
\end{array}
\end{eqnarray*}
\caption{The bird's eye view of the absolute value of the $i=T$ component 
of the wave functions of the Nambu-Goldstone fermions 
on the junction in the $(x^1,x^2)$ space
}
\label{FIG:NGf}
\end{figure}

We plot the absolute values of $|\psi_0^{(a) i=T}|$ of the $i=T$ component 
of the wave function of the Nambu-Goldstone fermions 
$a={\rm I}, {\rm II}, {\rm III}$ 
in Fig.~\ref{FIG:NGf}. 
We can see that Nambu-Goldstone fermions have wave 
functions which extend to infinity along three walls. 
They become identical to fermion zero modes on at least two of the walls 
asymptotically 
and hence they are not localized around the center of the junction. 
We can construct a linear combination of the Nambu-Goldstone fermions 
to have no support along one out of the three walls. 
However, no linear combination of these Nambu-Goldstone fermions 
can be formed which does not have support extended along any of the 
wall. 
Therefore these wave functions are not localized and are not normalizable. 
This fact means that 
the low energy dynamics of BPS junction cannot be described by a $1+1$
dimensional effective field theory with a discrete particle spectrum. 

Similarly the Nambu-Goldstone bosons corresponding to the broken translation 
$P^a, a=1, 2$ are given by  
\begin{equation}
A_0^{(a)}(x^1, x^2) = \partial_a A_{\rm cl}^i(x^1, x^2), \qquad a=1, 2. 
\end{equation}
These two bosonic massless modes consist of two left-moving modes and 
two right-moving modes. 
On the other hand, we have seen already that there are two left-moving 
massless Nambu-Goldstone fermions and one right-moving massless 
Nambu-Goldstone fermion. 
These two left-moving Nambu-Goldstone bosons and fermions form 
two doublets of the $(1,0)$ supersymmetry algebra. 
The right-moving modes are asymmetric in bosons and fermions: 
two Nambu-Goldstone bosons and a single Nambu-Goldstone fermion. 
These three states are all singlets of the $(1,0)$ supersymmetry algebra 
in accordance with our analysis in sect.\ref{sc:unitaryrep}. 
Therefore we obtained a chiral structure of Nambu-Goldstone fermions 
on the junction background configuration. 

\subsection{
Non-normalizability of the Nambu-Goldstone 
fermions 
}

We would like to argue that our observation is a generic feature 
of the Nambu-Goldstone fermions on the domain wall junction in a flat 
space in the bulk: Nambu-Goldstone fermions 
are not localized 
at the junction and hence are not normalizable, 
if they are associated with the supersymmetry 
breaking due to the coexistence of nonparallel domain walls. 
The following observation is behind this assertion. 
A single domain wall breaks only a half of supercharges. 
Nonparallel wall also breaks half of supercharges, some of which 
may be linear combinations of the supercharges already broken by the 
first wall. 
If the junction configuration is a $1/4$ BPS state, 
linearly independent ones among these two sets of broken supercharges 
of nonparallel walls become ${3 \over 4}$ of the original supercharges. 

To see in more detail, let us first note that the junction configuration 
reduces asymptotically to a wall if one goes along the wall, say the 
wall $1$. Let us denote the number of original supercharges to be $N$.
On the first wall, a half of the original supersymmetry 
($Q^{(1)}, \cdots, Q^{(N)}$) is broken. We call 
these broken supercharges as $Q^{(1)}, \cdots, Q^{(N/2)}$. 
Consequently we have Nambu-Goldstone fermions localized around 
the core of the wall and is constant along the wall. 
In the junction configuration, we have other walls which are not parallel 
to the first wall. 
Asymptotically far away along one of such walls, say wall $2$, 
another half of the 
supersymmetry  $Q'^{(1)}, \cdots, Q'^{(N/2)}$ is broken. 
If the junction is a $1/4$ BPS state, a half of these, 
say $Q'^{(1)}, \cdots, Q'^{(N/4)}$, is a linear combination of  
$Q^{(1)}, \cdots, Q^{(N/2)}$ broken already on the wall $1$. 
The other half,  $Q'^{({N \over 4}+1)}, \cdots, Q'^{({N\over 2})}$ 
are unbroken on the wall $1$. 
Altogether a quarter of the original supercharges remain unbroken. 
Consequently the Nambu-Goldstone fermions corresponding to 
 $Q'^{(1)}, \cdots, Q'^{(N/4)}$ have a wave function which extends to infinity 
and approaches a constant profile along both the walls $1$ and $2$. 
Those modes corresponding to 
 $Q'^{({N \over 4}+1)}, \cdots, Q'^{({N\over 2})}$ 
have support only along the wall $2$, and those corresponding to 
the linear combinations of $Q^{(1)}, \cdots, Q^{(N/2)}$  
orthogonal to $Q'^{(1)}, \cdots, Q'^{(N/4)}$ have support only along the wall 
$1$. 
Thus we find that any linear combinations of the Nambu-Goldstone fermions 
have to be infinitely extended along at least one of the walls which form 
the junction configuration. 
Therefore the Nambu-Goldstone fermions associated with the coexistence 
of nonparallel domain walls are not localized at the junction and 
are not normalizable. 

In our exact solution, domain wall junction configuration reduces 
asymptotically to the wall $1$ at $x^2 \to -\infty$ with fixed $x^1$. 
On the wall, only two supercharges in Eqs.(\ref{eq:brokencharges1})--
(\ref{eq:brokencharges}) are broken 
\begin{equation}
 Q_{\rm I} = \frac{1}{\sqrt{2}}
              (e^{i\pi/4}Q_2 + e^{-i\pi/4}\overline{Q}_{\dot{2}}), 
\end{equation}
\begin{equation}
 Q_{\rm II} = \frac{1}{\sqrt{2}}
              (e^{-i\pi/4}Q_1 + e^{i\pi/4}\overline{Q}_{\dot{1}}) ,
\end{equation}
and there are two corresponding Nambu-Goldstone fermions which become 
domain wall zero modes asymptotically 
\begin{eqnarray}
    \psi_0^{({\rm I})i}(x^1, x^2)&\!\!\!=&\!\!\!
    \left( 
    \begin{array}{c}
      4\partial_z A_{\rm cl}^i(x^1,x^2) e^{-i\pi/4} \\
       0 
    \end{array}
    \right) 
    \\
  &\!\!\!  \rightarrow &\!\!\!
    \left( 
    \begin{array}{c}
      2\partial_1 A_{\rm cl}^{i{\rm wall}}(x^1) e^{-i\pi/4} \\
       0 
    \end{array}
    \right), \nonumber \\
    \psi_0^{({\rm II})i}(x^1,x^2)&\!\!\!=&\!\!\!
    \left( 
    \begin{array}{c}
       0 \\  
       2\partial_1 A_{\rm cl}^i(x^1,x^2) e^{i\pi/4}
    \end{array}
    \right)\nonumber \\
 &\!\!\!   \rightarrow &\!\!\!
    \left( 
    \begin{array}{c}
       0 \\  
       2\partial_1 A_{\rm cl}^{i{\rm wall}}(x^1) e^{i\pi/4}
    \end{array}
    \right).
\label{NG1-1}
\end{eqnarray}
These wave functions are localized on the core of the wall 1 
in the $x^1$ direction and are constant along the wall. 
Along the other walls we find two broken supercharges one of which 
is identical to one of the broken supercharges, $Q_{\rm I}$. 
The other broken supercharge is $Q'_{\rm II}$ on the wall 2 
and  $Q''_{\rm II}$ on the wall $3$. 
There are only two independent supercharges among 
 $Q_{\rm II}$,  $Q'_{\rm II}$, and  $Q''_{\rm II}$. 
Together with  $Q_{\rm I}$ we obtain three independent broken supercharges. 
We can construct a linear combination of the Nambu-Goldstone fermions to 
have no support along one out of the three walls. 
However, any linear combination has nonvanishing wave function which becomes 
fermion zero mode on at least one of the wall asymptotically. 
Therefore the associated Nambu-Goldstone fermions have support 
which is infinitely extended at least along two of the walls. 

If a single wall is present, we can explicitly construct 
a plane wave solution propagating along the wall, 
which may be called a spin wave and is among massive modes on the wall 
background. 
Even if there are several walls forming a junction configuration, 
we can consider excitation modes which reduce to the spin wave modes 
along each wall. 
They should be a massive mode on the domain wall junction background. 
The Nambu-Goldstone mode on the domain wall junction is 
the zero wave number limit of such a spin wave mode. 
This physical consideration suggests that the massless Nambu-Goldstone 
fermion is precisely the vanishing wave number (along the wall) 
limit of the massive spin wave mode. 

Let us note that our argument does not apply to 
models with the bulk cosmological constant.
In such models, massless graviton is localized on the background
of intersection of walls\cite{ADDK}.
In that case, massless mode is a distinct mode different from the massless 
limit of the massive continuum, although the massless mode is buried 
at the tip of the continuum of massive modes. 
The normalizability of the massless graviton is guaranteed by 
the Anti de Sitter geometry away from the junction or intersection 
including the direction along the wall.

\subsection{Negative contribution of central cha- \\
rge $Y_3$ to junction mass}
Next we discuss the sign of the contribution of the 
central charge $Y_3$ to the mass of the junction configuration. 
We can use the Stokes theorem to obtain an expression for 
the central charge $Y_3$ as a contour integral\cite{OINS},\cite{CHT} 
\begin{eqnarray}
Y_3 &\!\!\!=&\!\!\!
\int d x^3 \,
i  
\int d^2 x \,
\left[\partial_1\left(K_{i}
 \partial_2 A^i\right)
-\partial_2\left(K_{i}
 \partial_1 A^i\right)
\right] \nonumber \\
&\!\!\!=&\!\!\!
\int d x^3 \,
i  
\oint K_{i} d A^i, 
\label{y-area-generic}
\end{eqnarray}
where $K_i \equiv \partial K / \partial A^i$ is a derivative 
of the K\"ahler potential $K$. 
This contour integral 
in the field space should be done as a map from a counterclockwise 
contour in the infinity of $z=x^1+ix^2$ plane. 
Only complex fields can contribute to $Y_3$. 
Let us assume for simplicity that there is only one field which can 
contribute to $Y_3$ as in our exact solution. 

Eq.(\ref{y-area-generic}) shows that 
the central charge $Y_3$ becomes negative (positive), 
if the asymptotic counterclockwise contour in $x^1, x^2$ is mapped 
into a counterclockwise (clockwise) contour in field space. 
On the other hand, the sign of the contribution of the central charge 
$Y_3$ to the mass of the junction configurtion is determined by 
the formula, 
$H=H_{\rm II} = |\langle i Z_1-Z_2 \rangle| +\langle Y_3 \rangle $ 
or $H=$
$H_{{\rm I}}=|\langle -i Z_1-Z_2 \rangle| -\langle Y_3 \rangle $. 
The choice of these mass formulas are in turn deterined by the map 
of the asymptotic counterclockwise contour in $x^1, x^2$ space to a 
 counterclockwise or clockwise contour in the superpotential 
space ${\cal W}$. 
Combining these two observations, we conclude that the contribution of the 
central charge $Y_3$ to the mass of the junction configuration is negative 
if the sign of rotations is the same in field space $A^i$ and in 
superpotential space ${\cal W}$, and positive if the sign of rotations is 
opposite.

The field configuration moves counterclockwise in field space 
in our exact solution in (\ref{exact solution}) and then 
the central charge is negative in this solution. 
Since the exact solution satisfies the BPS equation 
for the case $H=H_{\rm II}$, 
 the central charge contributes to the mass of the junction 
configuration negatively. 
Therefore we should not consider the central charge $Y_3$ alone 
as the physical mass of the junction at the center. 
In the junction configuration,  the junction at the  center 
cannot be separated from the walls. 
We also can find a solution for the other case of $H=H_{\rm I}$ 
in our model. 
The solution is just a configuration obtained by a reflection 
$x^1 \rightarrow -x^1$. 
Then the central charge is positive, but the contribution to 
the mass $H=H_{\rm I}$ becomes again negative.  
In either solution, the rotation in field $T$ space has the same sign 
as the rotation in superpotential ${\cal W}$ space. 
Therefore central charge $Y_3$ contributes negatively to the mass 
of the junction, irrespective of the choice of 
$H=H_{{\rm I}}$ or $H=H_{{\rm II}}$.  

More recently this feature of negative contribution of 
 $Y_3$ to the junction mass is studied from a different viewpoint and 
it is argued that this feature is valid in most situations 
except possibly in contrived models\cite{ShifmanVeldhuis}. 

\section{Summary}
\begin{enumerate}
\item 
 We have obtained an exact solution of domain wall junction in a 
four-dimensional ${\cal N}=1$ SUSY $U(1)\times U(1)'$ gauge 
theory. 
 The model has three pairs of chiral superfields and is motivated by the 
${\cal N}=2$ $SU(2)$ gauge theory with one flavor perturbed by an adjoint 
scalar mass. 
\item 
Mode functions are defined and are shown to appear in a boson-fermion pair of 
identical mass for massive modes. 
\item
Nambu-Goldstone fermions exhibit a chiral structure in accordance with 
($1, 0$) supersymmetry. 
\item
Nambu-Goldstone fermions are not normalizable. 
The domain wall junction configuration preserves only $1/4$ of the original 
supercharges, 
but modes on the junction does not reduce to 
a $1+1$ dimensional field theory with a discrete mass spectrum 
even for massless modes. 
\item We find that the new central charge $Y_k$ associated with the junction 
gives a negative contribution to the mass of the domain wall junction, 
whereas the central charge $Z_k$ gives a dominant positive contribution. 
One has to be cautious to identify 
the central charge $Y_k$ alone as the mass of the junction. 
\end{enumerate}

\end{document}